# Observation of 0 to π transition in YBCO/Gallium oxide multilayers

L.S. Vaidhyanathan, M.P. Janawadkar and D.K Baisnab

Condensed Matter Physics Division,
Materials Science Group
Indira Gandhi Centre for Atomic Research, Kalpakkam 603 102
Tamil Nadu, India

L.S.Vaidhyanathan (lsv@igcar.gov.in)

M.P. Janawadkar (mpj@igcar.gov.in)

D.K. Baisnab (dkbb@igcar.gov.in)

For all authors: Tel/Fax: 0091-44-27480081

When two superconductors are coupled via an insulator the phase difference between them is expected to be zero in the ground state and such Josephson junctions are known as "0" type Josephson junctions . But in the presence of magnetic impurities in the barrier or ferromagnetism, the phase difference in the ground state could become $\pi$; such Josephson junctions are known as "$\pi$" type Josephson junctions[1-4]. Experimental results have confirmed the observation of a 0 to $\pi$ transition as a function of temperature in Josephson junctions such as Nb/Cu$_{1-x}$Ni$_x$/Nb based on ferromagnetic weak links with low exchange energy[5]. Recently, observation of possible magnetic behaviour in oxides has been reported[6,7], and it would, therefore, be interesting to study the Josephson coupling in YBCO/gallium oxide multilayers. This work reports the observation of a very anomalous superconducting behaviour in YBCO/gallium oxide multilayers. A transition to the superconducting state has been observed with an enhanced superconducting transition temperature of 98 K in the YBCO/gallium oxide multilayer followed by a second transition temperature at 90 K with the multilayer losing its superconductivity at intermediate temperatures.  This behaviour is explained by assuming that the Josephson coupling between the superconducting YBCO layers in the multilayer changes from $\pi$ type to 0 type. This is the first experimental observation of the 0 to $\pi$ transition in HTSC/gallium oxide/HTSC thin film structures.

Recent research has revealed a rich variety of cooperative phenomena involving superconducting and magnetic orders. Although the phenomenon of reentrant superconductivity has been known in the context of superconductors with Kondo impurities[8] and ferromagnetic superconductors[9,10], experimental observation of a reentrance into the superconducting state has remained elusive so far. In the thin film form, when a conventional spin singlet superconductor is brought into contact with a ferromagnetic thin film, the Cooper pair correlations extend over a length of only a few nanometers across the S-F interface. The exchange field in the ferromagnet creates a spatially varying phase which results in an oscillatory variation of the critical current as a function of ferromagnetic layer thickness[1,11,12]. This leads to a number of oscillatory phenomena in S/F systems, including the oscillations in the superconducting transition temperature of S/F multilayers[13].

A Josephson junction based on Superconductor-Insulator-Superconductor (SIS) sandwich structure is usually described by the phase difference $\phi$ between the two superconductors and has a current-phase relationship given by $I = I_c \sin\phi$ with a ground state corresponding to $\phi = 0$. When magnetic impurities are present in the insulator layer of SIS, it was proposed theoretically[14] that, in certain regime of parameters, the spin flip scattering in the barrier may favour a state corresponding to $\phi = \pi$ and the ground state of the superconducting ring incorporating such a junction will have a non-zero value for the enclosed magnetic flux. In recent years, attempts have been made to realize the "$\pi$" type Josephson junctions[15]. Measurement of temperature dependence of critical current in some junctions has indicated an anomalous behaviour including a sharp cusp as the junction is cooled below the $T_c$ of the superconducting electrodes; indeed, this observation was taken to signify the manifestation of a change from "0" to "π" type Josephson junction[15]. The superconductor-ferromagnet-superconductor (SFS) Josephson junction could potentially exhibit a "π" type behaviour due to oscillation of the order parameter in the vicinity of the ferromagnetic tunnel barrier. Experimental studies carried out on these systems show oscillation in superconducting transition temperature $T_c$ with ferromagnetic layer thickness[13]; the oscillations have been interpreted as arising due to stabilisation, in certain ranges of thickness, of a higher $T_c$ π phase[16]. Initial experiments performed on S/F/S multilayers involved the use of ferromagnetic materials with a high Curie temperature[17]. In general, non-monotonic variation of $T_c$ was observed[13] and attributed either to the π state[16] or to the onset of ferromagnetism[18]. The superconducting order parameter decays over a length scale $\xi_F$ as it penetrates into the F-layer of the SFS sandwich structure. $\xi_F$ depends on the exchange energy $E_{exch}$ in the ferromagnet. An estimate of $\xi_F \sim (\hbar D/E_{exch})^{0.5}$ yields a value of the order of 1 nm in case of ferromagnetic layers such as Fe, Co and Ni. In view of this, research shifted to investigations on S/F/S systems based on weak ferromagnets with a low value of Curie temperature. Ryazanov et al[1] observed a nonmonotonic temperature dependence of the critical current of SFS Josephson junctions employing a weakly ferromagnetic alloy such as $Cu_xNi_{1-x}$ as a ferromagnetic layer and by studying the temperature dependence of the critical current, provided a possible evidence for the existence of the π state. Several recent experimental investigations have served to confirm the existence of π state in such structures[19].

In this context, it becomes interesting to investigate the establishment of phase coherence in multilayers comprising of superconducting layers separated by thin films of oxides which are non-magnetic in bulk but are ferromagnetic in nano-form. Possibility of magnetism in several nonmagnetic oxides is being actively investigated in the current literature[6,7] with a view to understand the origin of magnetism in such oxides and its influence on the physical properties.

In this work, we report an investigation on the superconducting behaviour of YBCO/gallium oxide multilayers with the thickness of the gallium oxide layer maintained at very small values (~ 2nm). Our experimental observations reveal an extremely anomalous superconducting behaviour in this system.

Thin films of gallium oxide having a thickness of 100 nm were deposited over various types of substrates like Si and quartz by pulsed laser deposition technique using KrF laser by varying deposition parameters such as laser energy, oxygen pressure and temperature of deposition to optimize the deposition process. Structural characterization was performed through glancing incidence X-ray diffraction (GIXRD) at room temperature for films deposited at a substrate temperature of 800 C with laser energy of 200 mJ under oxygen gas pressure of 7 Pa, and indicated the formation of γ-phase of bulk nanocrystalline gallium oxide (Fig.1.)

The multilayer thin films of YBa$_2$Cu$_3$O$_{7-\delta}$ (YBCO) and gallium oxide were then deposited by Pulsed Laser Deposition (PLD) on MgO (100) substrates with KrF excimer laser with a view to see if thin layers of gallium oxide affect the superconducting properties of YBCO. Single layer YBCO and gallium oxide thin films were initially characterized for their thickness using a Dektak 3030A surface profiler. During deposition, the substrate temperature was maintained at 800C in flowing oxygen atmosphere. YBCO films were deposited at oxygen pressure of 50 Pa while the gallium oxide films were deposited at a lower oxygen pressure of 7 Pa. Films were deposited in the constant energy mode with an energy per pulse of 200 mJ. Each multilayer had 10 layers of YBCO and gallium oxide. The films were subsequently annealed in oxygen at a pressure of ~ 9.5 x 10$^4$ Pa and were cooled down to room temperature with different cooling rates but with a 45 minutes stay at 550 C and 500 C. YBCO layer thickness was varied between 25 and 60nm, while gallium oxide layer thickness was maintained at ~ 2nm. Transport measurements were performed down to liquid helium temperatures by four-probe contact technique using indium soldering to establish the contacts.

Fig.2 shows the normalized electrical resistance as a function of temperature for these multilayers along with that for a single layer YBCO thin film deposited on MgO (100) substrate for comparison. The multilayer YBCO/gallium oxide (25 nm/2 nm) was found to have a T$_c$ slightly lower than that observed for other multilayers. In fact, a slight enhancement of T$_c$ was seen for other multilayers in comparison to the single layer YBCO thin film. We now present a detailed discussion of our experimental observations for a selected sample of YBCO/ gallium oxide multilayer (60 nm/2 nm).

Fig. 3 shows the variation of the normalized resistance as a function of temperature for the YBCO/ gallium oxide multilayer (60nm / 2nm) showing a reentrant superconducting behaviour. As the multilayer was cooled, the onset of superconductivity was seen at a temperature of 98K. When cooled further, however, resistance was restored at a lower temperature of 93.7K. Finally, resistance started to decrease again at a temperature of 90.1K with the appearance of zero resistance state at temperatures below 81K. As shown in fig.3, hysteresis was observed between the measurements carried out during the warm-up and cool-down cycles but the reentrant behaviour was evident during both the cycles.

Ryazonov et al[1] have studied Josephson junctions based on thin layer of a weak ferromagnet. Investigations on the temperature dependence of the critical current $I_c$ in Nb/Cu$_{0.48}$Ni$_{0.52}$/Nb junction showed that $I_c$ increases with decrease in temperature, goes through a maximum, decreases to zero with a sharp cusp at the node and then rises again sharply with further decrease in temperature. This anomalous temperature dependence of $I_c$ has been rationalized in terms of a change in Josephson coupling from "0" to "π" type.

In the case of cross junctions based on dilute ferromagnetic alloys such as Nb-CuNi-Nb[20], experimental investigations on the temperature dependence of $I_c$ have shown a reentrant superconducting behaviour for junctions with 22 nm thick Cu-52at%Ni film having a ferromagnetic Curie temperature as low as

20K. The authors conclude that the SFS junction has a tendency to transform from 0 to π state when the thickness and Curie temperature of the ferromagnetic layer are suitably adjusted.

To explain our results, we consider the model proposed by Shu Yamaguchi[21] in which gallium oxide is modeled as a mixture of phases viz crystalline gallium oxide dispersed in a matrix of amorphous gallium oxide; this gives rise to an excess of gallium with concomitantly produced oxygen vacancies and/or Ga interstitials. These defects act as electron donors and the present work focuses on what would happen to the superconducting properties of YBCO if it is in the neighbourhood of such gallium oxide thin films.

If the phase difference between the two superconducting layers separated by gallium oxide happens to be π, we speculate that this can result in the observation of a higher $T_c$ in the multilayer and the observation of $T_c$ onset of 98 K in this sample could be due to the phase coupling of π type. We also see a second transition temperature at a lower temperature of 90 K and this could be ascribed to the phase coupling of 0 type. Between these two temperatures, critical current may decrease to zero, thereby generating the observed resistance at intermediate temperatures. We have not observed this type of 0 to π transition for other samples prepared during this study. We think that such a situation could be relevant even in the context of SFS junctions. In SFS junctions, when the barrier thickness was 1 to 2 nm away from the nodes, 0 to π transition was not observed although the same magnetic layer was used.

It may be noted that this is the first experimental report of the observation of a 0 to π transition as a function of temperature in YBCO/gallium oxide multilayers although Shiro Kawabata et al predicted the possibility of 0 to π state transition in HTSC/Ferromagnetic insulator/HTSC junctions[22] some years ago.

Acknowledgements:

We thank Dr V. Sridharan for help in preparation of Gallium oxide sample.

Author Contributions:

All authors contributed substantially to this work

Author Information:

Reprints and permissions information is available at www.nature.com/reprints. The authors declare no competing financial interests. Correspondence and requests for materials should be addressed to LSV ( lsv@igcar.gov.in )

Figure legends:

**Figure 1** Glancing angle X-ray diffraction pattern of gallium oxide thin films deposited by pulsed laser deposition on several substrates (a) on quartz (b) on Si and (c) bulk γ gallium oxide

**Figure 2** Temperature dependence of the resistance of different YBCO / gallium oxide multilayers. Note the reentrant transition observed for YBCO / gallium oxide (60/2) nm multilayer. Inset shows the superconducting transition temperature of YBCO thin film.

**Figure 3** Hysteresis observed around the reentrant transition for YBCO / gallium oxide (60/2) nm multilayer.

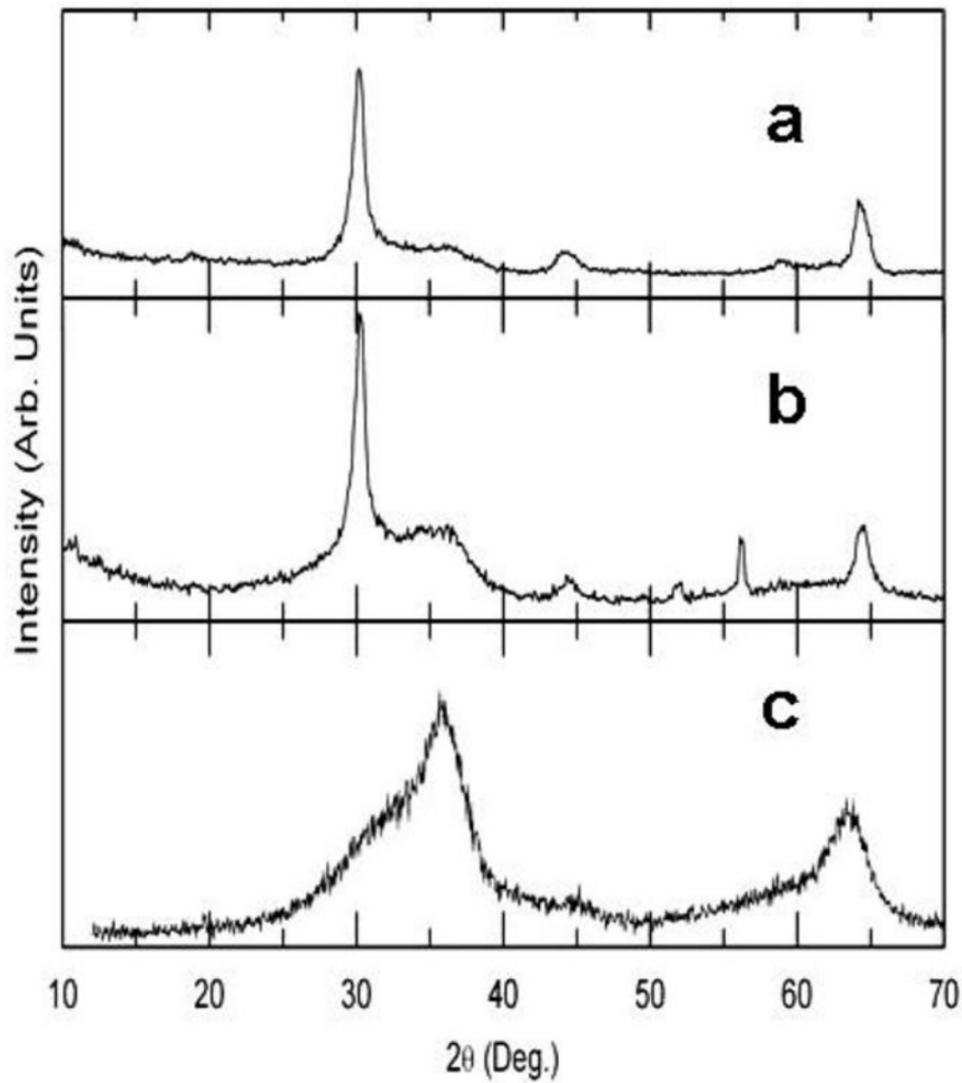

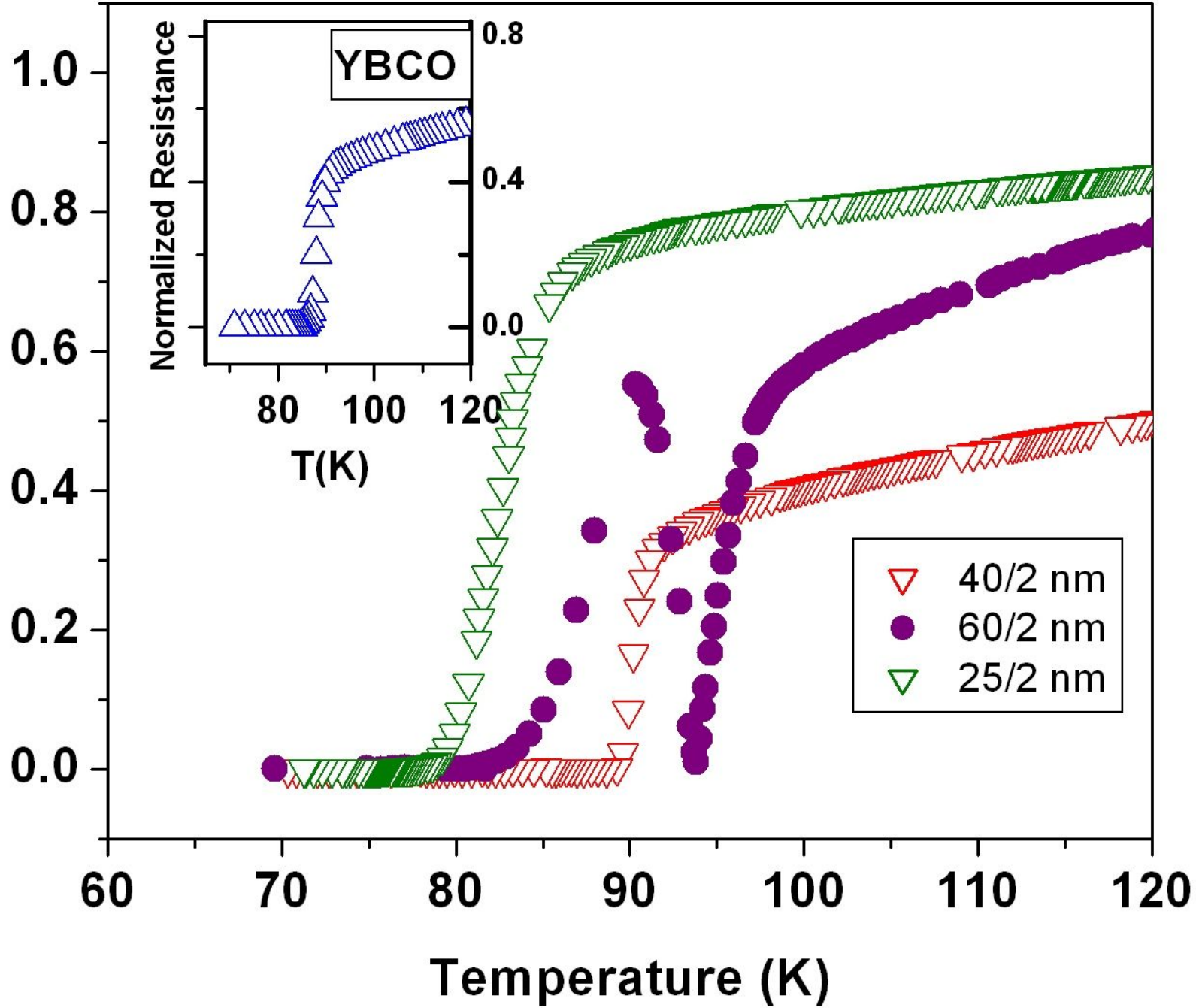

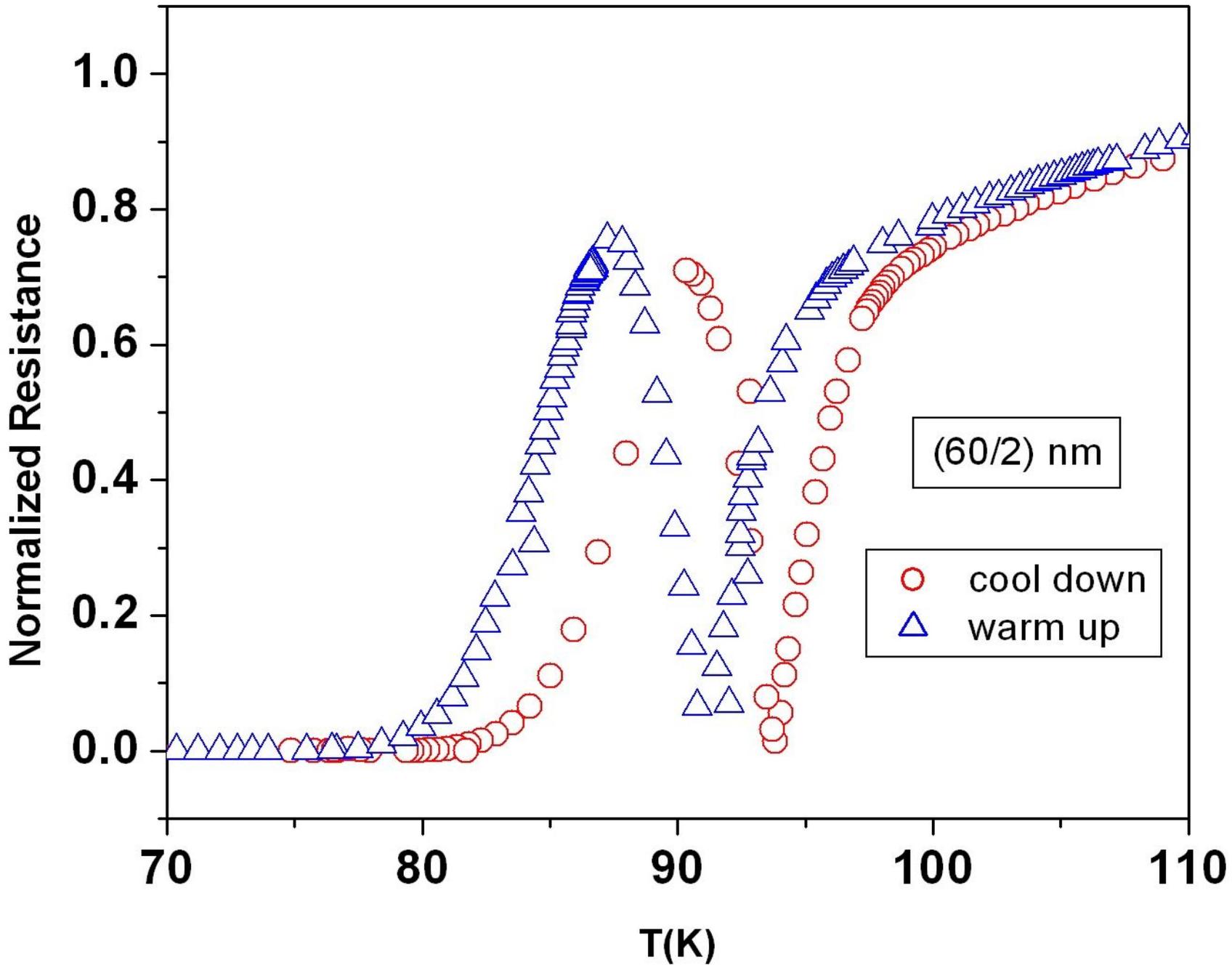